\documentclass{article}

\usepackage{graphicx, wrapfig, subcaption, setspace, booktabs}

\usepackage{arxiv}

\usepackage[utf8]{inputenc} % allow utf-8 input
\usepackage[T1]{fontenc}    % use 8-bit T1 fonts
\usepackage{hyperref}       % hyperlinks
\usepackage{url}            % simple URL typesetting
\usepackage{booktabs}       % professional-quality tables
\usepackage{amsfonts}       % blackboard math symbols
\usepackage{nicefrac}       % compact symbols for 1/2, etc.
\usepackage{microtype}      % microtypography
\usepackage{lipsum}

\title{Mapping the Hubbard model to the t-J model using ground state unitary transformations}

\author{
  Yifan Tian \\
  Department of Physics and Astronomy\\
  University of California Irvine\\
  Irvine, CA 92697 \\
  \texttt{yifant@uci.edu} \\
}

\begin{document}
\maketitle

\begin{abstract}
The effective low-energy models of the Hubbard model are usually derived from perturbation theory. Here we derive the effective model of the Hubbard model in spin space and t-J space using a unitary transformation from numerical optimization. We represent the Hamiltonian as Matrix product state(MPO) and represent the unitary transformation using gates according to tensor network methods. We obtain this unitary transformation by optimizing the unitary transformation between the ground state of the Hubbard model and the projection of the Hubbard model ground state into spin space and t-J space. The unitary transformation we get from numerical optimization yields effective models that are in line with perturbation theories. This numerical optimization method starting from ground state provides another approach to analyze effective low-energy models of strongly correlated electron systems.
\end{abstract}

% keywords can be removed

\section{Introduction}
% \lipsum[2]
The effective low-energy models of the Hubbard model are usually derived from perturbation theory. Including Canonical Transformations\cite{c1} and Brillouin-Wigner Method etc\cite{c2}. Here we derive the effective low-energy Hamiltonian of the Hubbard model in spin space for half-filling and t-J space for doping cases using a unitary transformation by numerical optimization. 
The Hubbard model is defined as
 \begin{equation}
 H = -t \sum_{<i,j>\sigma} c^+_{i,\sigma}c^-_{i,\sigma} +h.c + U\sum_{<i,j>\sigma}n_{i,\sigma}n_{j,\sigma} 
 \end{equation}
The Effective Hamiltonian of the Hubbard model in spin space is defined as
 \begin{equation}
 H_{eff} = \sum_{<i,j>}(J*S_iS_j-a)
  \end{equation}
 If we write Hubbard H as different pieces as
 \begin{equation} 
 H = T_0 + T_1 + T_{-1} + V
 \end{equation}
 
where $T_i$ refers to one piece of the kinetic energy operator that change the of the doubly-occupied sites respectively $(+1,0,-1)$. 
Then Effective H in t-J space can be expressed as \cite{c3}

\begin{equation}
H_{eff} = V + aT_0 + bT_{-1}T_{1} + cT_{-1}T_{0}T_{1}+...
 \end{equation}

\section{Derivation of the unitary}
% \label{sec:headings}

There is an infinite number of unitaries that remove all double occupancy in the ground state. We choose the unitary that has the minimum number of gates. 
Hamiltonian is represented as a Matrix product operator (MPO). We obtain this unitary transformation by optimizing the unitary transformation that makes the norm of the ground state of the Hubbard model and the projection of Hubbard ground state into spin space or t-J space to be 1. So the loss function is defined as $1-\left\langle\psi(0) |psi \right\rangle $, where $\left| \psi(0) \right\rangle$ is the ground state of Hubbard model, $\left| \psi \right\rangle$ is the projection of Hubbard ground state into spin space or t-J space. The ground state is computed using DMRG method\cite{c4}. Figure 1 illustrates the approach. Since this unitary is derived from optimizing the norm between ground states and projected ground state, we call this method as ground state unitary transformation.

\begin{figure}
\centering
\begin{center} 
\includegraphics[width=0.6\columnwidth]{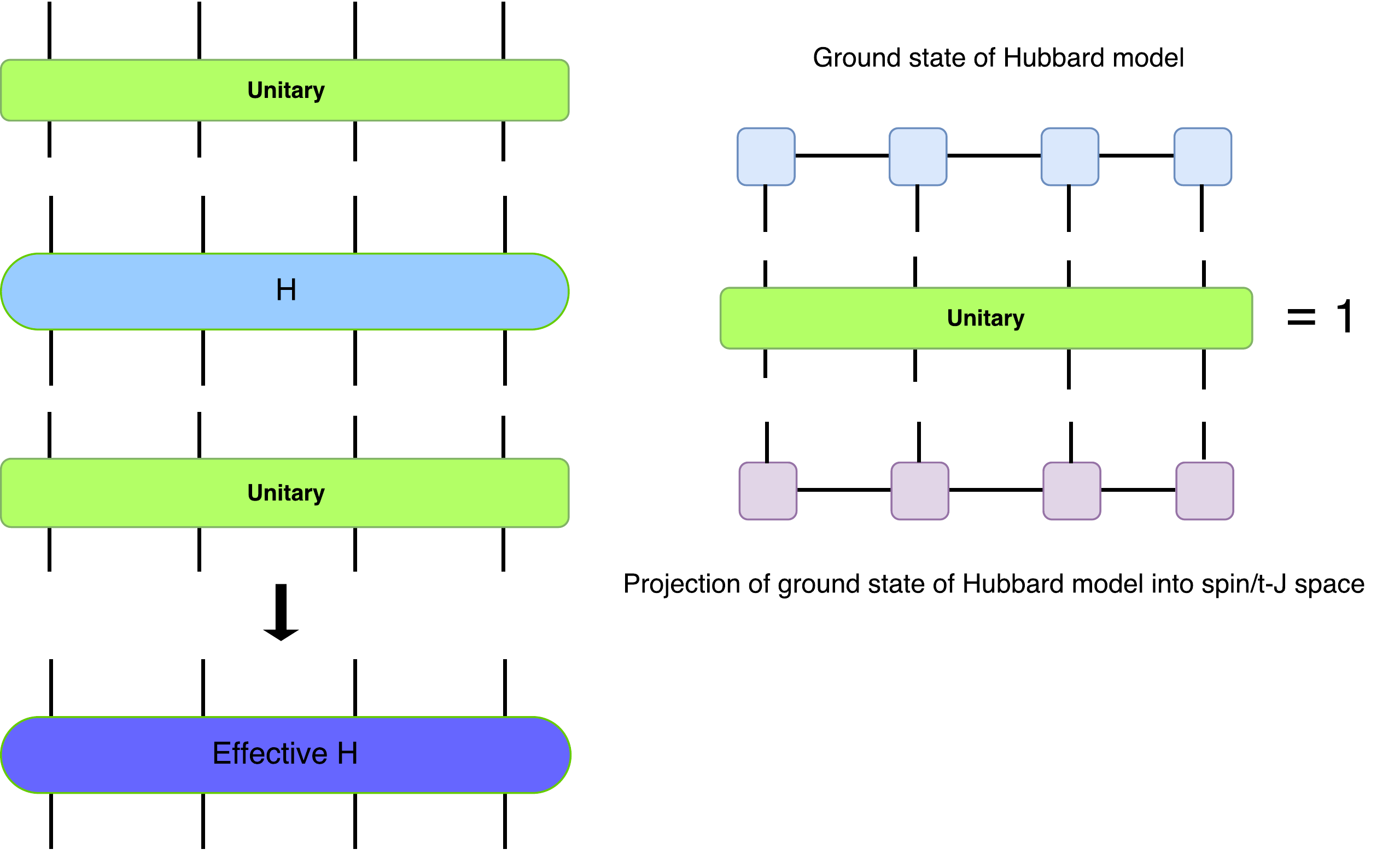} % Example image
\end{center}
\caption{The unitary transformation that project ground state of Hubbard mode into spin/t-J space(right). The same unitary can be applied to the Hamiltonian of the Hubbard model to generate the effective Hamiltonian in spin/t-J space.} 
\end{figure}

\section{Construction of the unitary}

The stack of gates can be used to construct unitary transformations. These gates are also called disentanglers in MERA algorithm\cite{c4}. We analyze two structures for constructing unitary using gates. The DMRG-like gates and the MERA-like gates. In half-filling case, the 3-sites DMRG-like gates (Left) can reach very high accuracy. Figure 3 shows the convergence of the gates versus the system-size and the convergence of the gates versus layers of gates for half-filling case. In half-filling case, the optimization can reach a very high accuracy using a three-sites DMRG-like gate.

\begin{figure}
\centering
\begin{subfigure}{.5\textwidth}
  \centering
  \includegraphics[width=.7\linewidth]{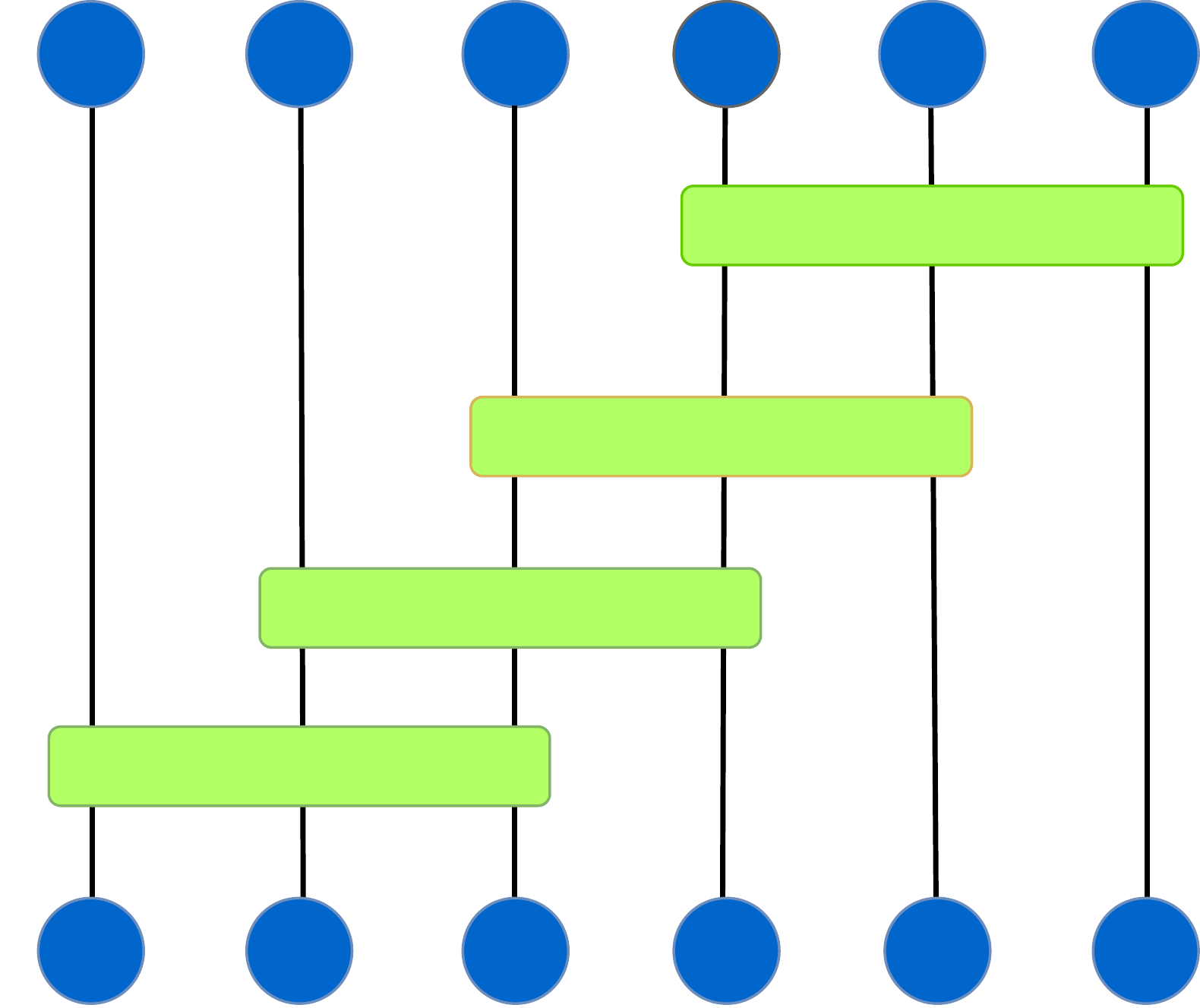}
%   \caption{A subfigure}
  \label{fig:sub1}
\end{subfigure}%
\begin{subfigure}{.5\textwidth}
  \centering
  \includegraphics[width=.7\linewidth]{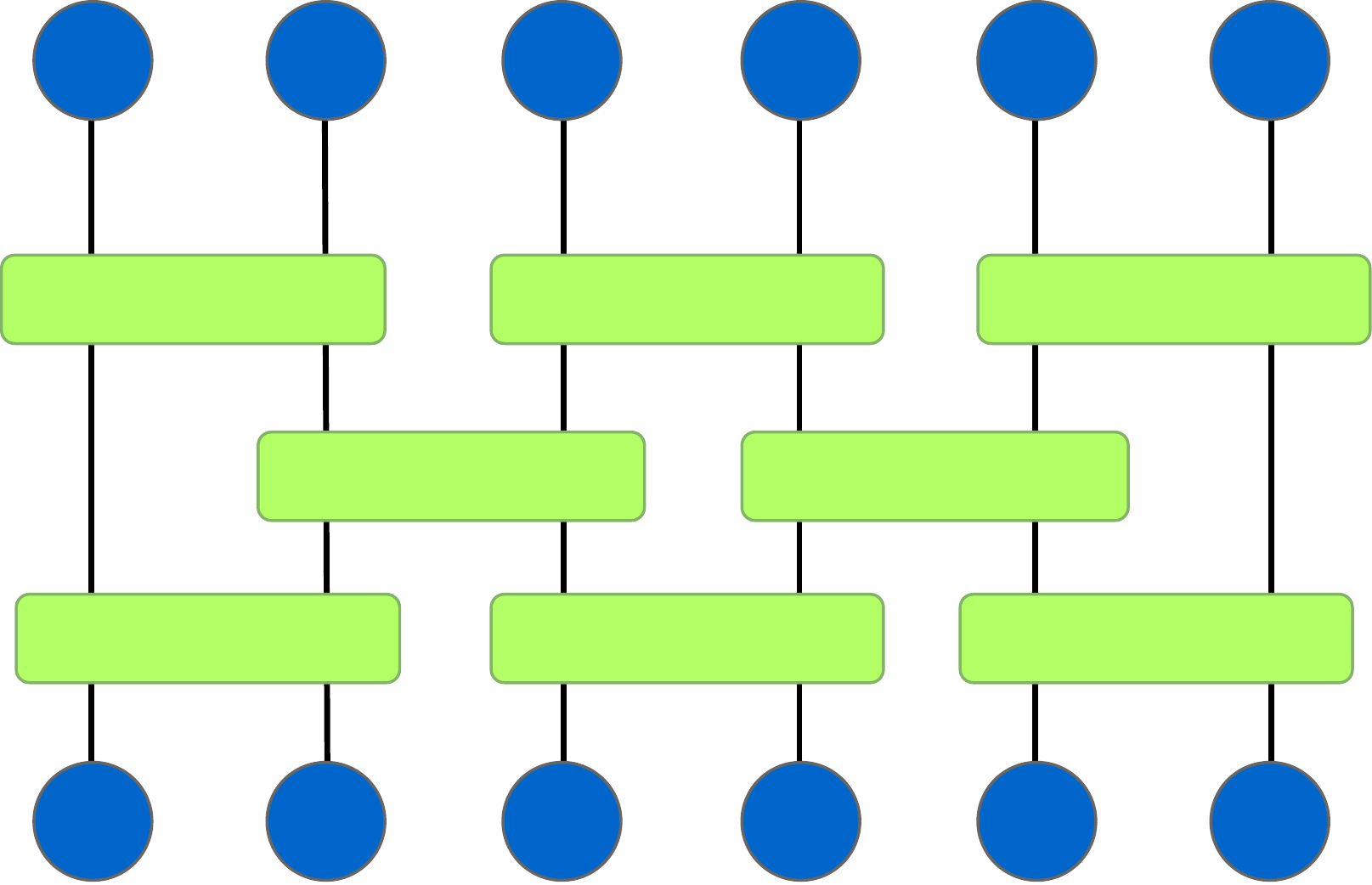}
%   \caption{A subfigure}
  \label{fig:sub2}
\end{subfigure}
\caption{Construction of unitary using two structures. The DMRG-like gates(Left) and the MERA-like gates(Right)}
\label{fig:test}
\end{figure}

\begin{figure}
\centering
\begin{subfigure}{.5\textwidth}
  \centering
  \includegraphics[width=.7\linewidth]{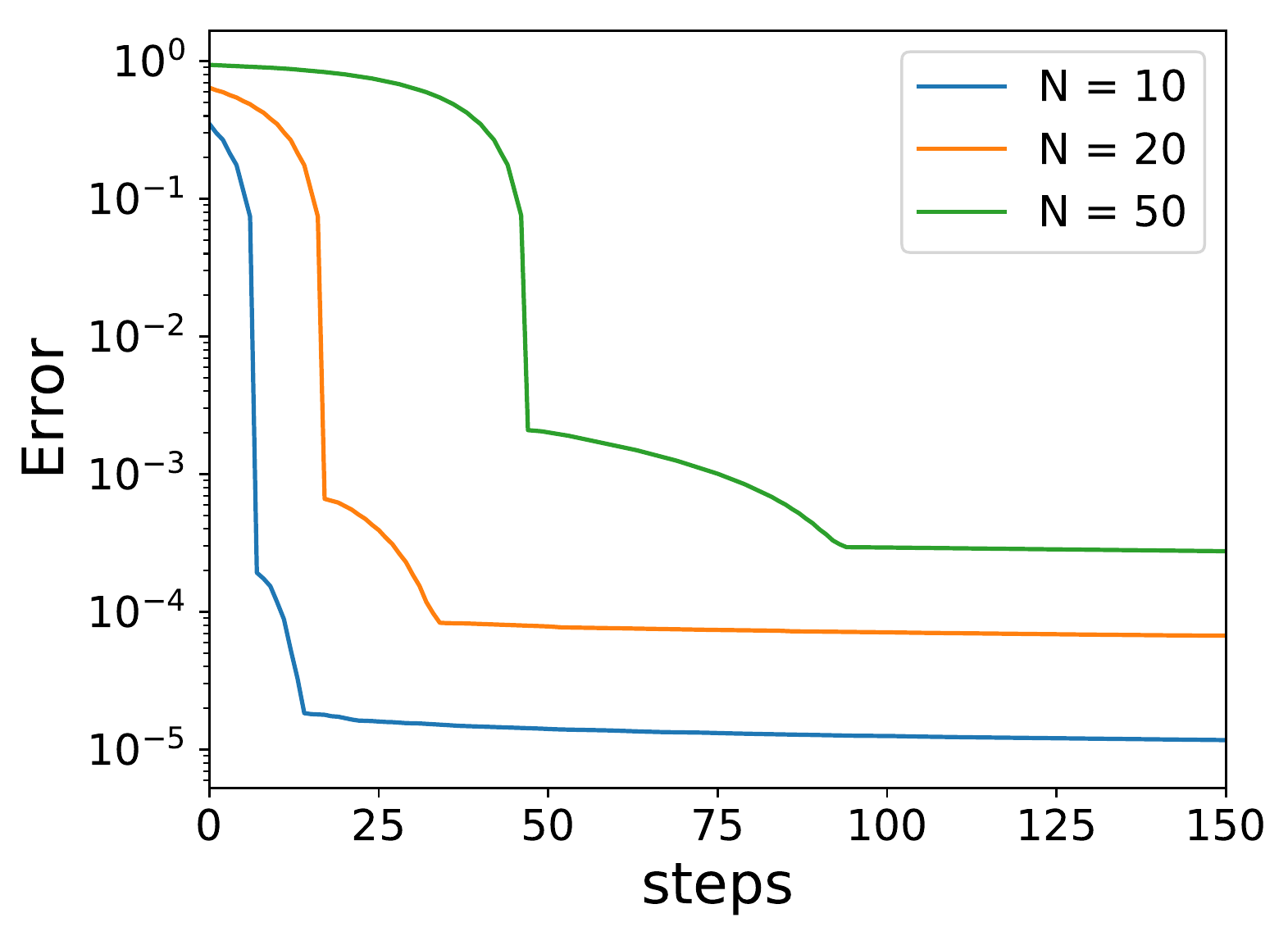}
%   \caption{A subfigure}
  \label{fig:sub1}
\end{subfigure}%
\begin{subfigure}{.5\textwidth}
  \centering
  \includegraphics[width=.65\linewidth]{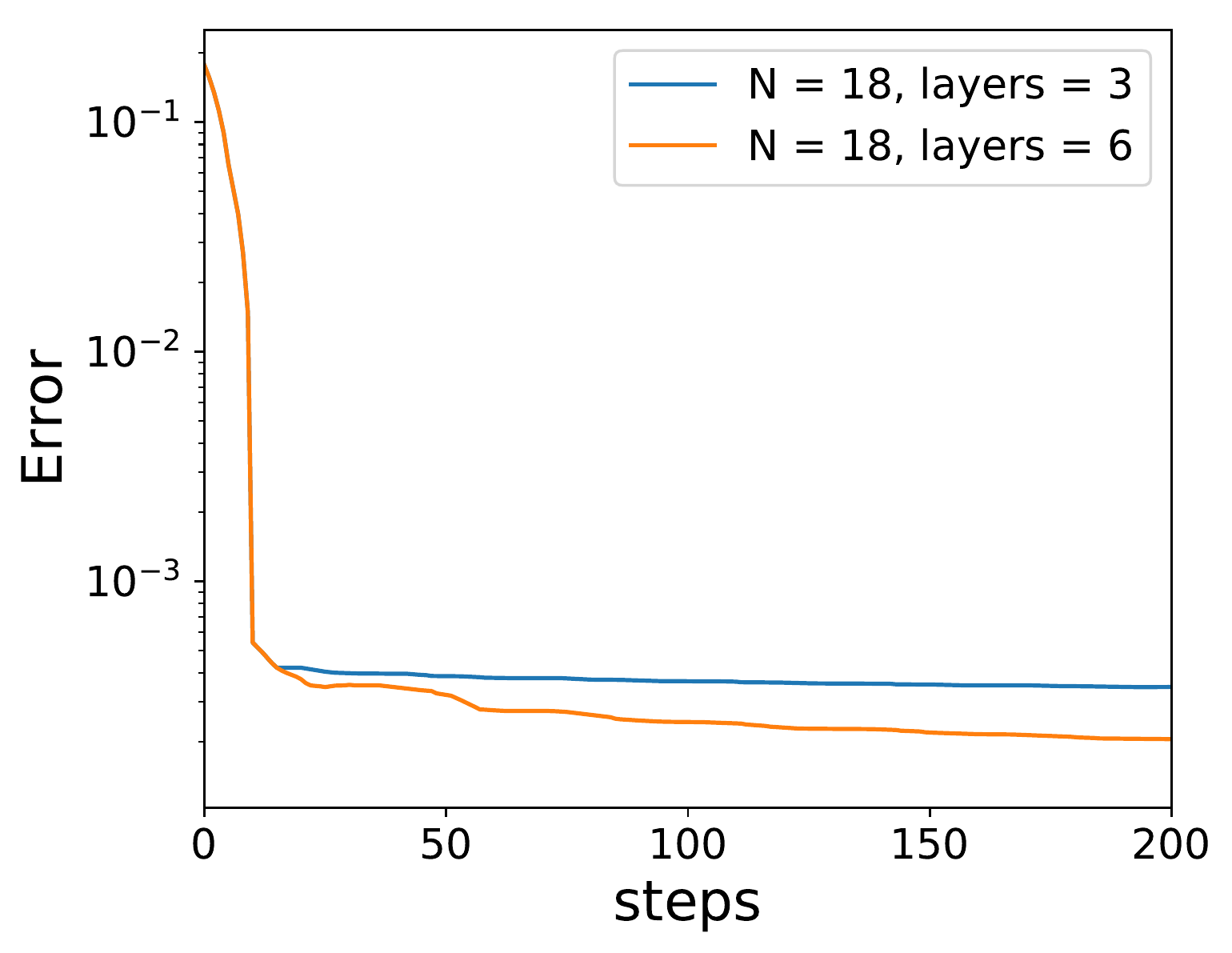}
%   \caption{A subfigure}
  \label{fig:sub2}
\end{subfigure}
\caption{The convergence of the error of the unitary versus the system size using DMRG-like gates(left). The convergence of the error of the unitary versus the number of layers using MERA-like gates(right). }
\label{fig:test}
\end{figure}

\section{Half-filling}

From perturbation theory. At half-filling case. When $U>>1$. There is a mapping between the Hubbard model and Heisenberg model\cite{c5}
\begin{equation}
J = \frac{4t^2}{U}, \   a = -\frac{1}{4} 
\end{equation}
After applying the unitary from optimization to the Hamiltonian. We get the effective Hamiltonian. We get the term J and term a by fitting the effective Hamiltonian yielded by the unitary transformation to the corresponding terms. Figure 4 shows the comparison between the fitting of the J from Hamiltonian yielded by the unitary transformation and the theoretical equation $J = 4*t^2/U$.

\begin{figure}
\centering
\begin{center} 
\includegraphics[width=0.5\columnwidth]{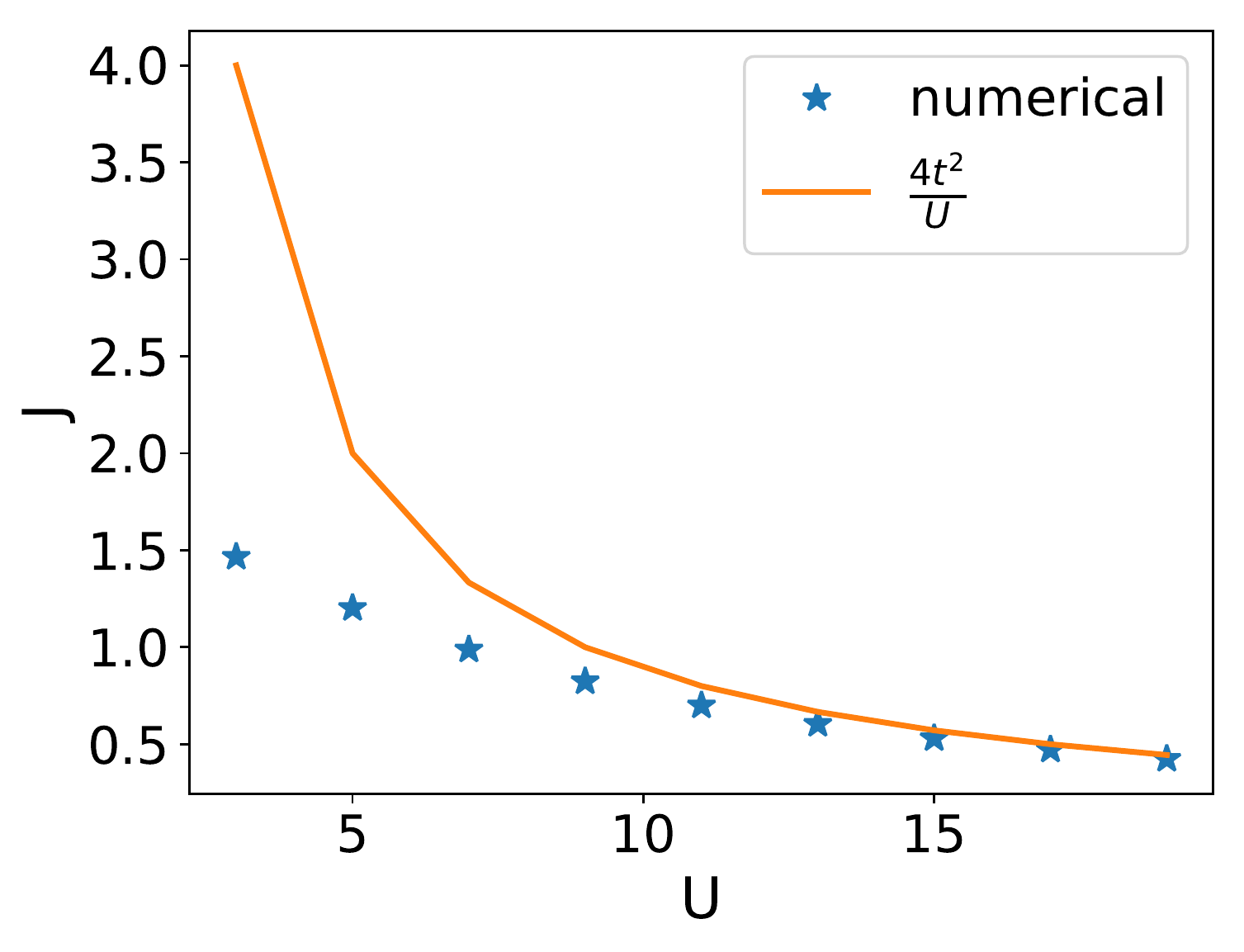} % Example image
\end{center}
\caption{The comparison between the fitting of the J from Hamiltonian yielded by the unitary transformation and the theoretical equation $J = 4*t^2/U$.} 
\end{figure}

\section{t-J doped results}
For doping cases. We construct Effective Hamiltonian using t-J space. In t-J space, the accuracy of the norm is much worse than the half-filling case. The cutoff we use to construct MPO is much higher as well. To achieve a better model. We have to use smaller cutoff in applying the unitary gates, which yield MPO with big bond dimension. The bottom left figure shows the comparison between low energy levels of Effective H and the Hubbard model. System size(N) = 6. Number of particles(P) = 4, cutoff=$10^{-3}$. 

% The right figure shows the ground state energy of effective H using different cutoff during construction.

\begin{figure}
\centering
\begin{center} 
\includegraphics[width=0.5\columnwidth]{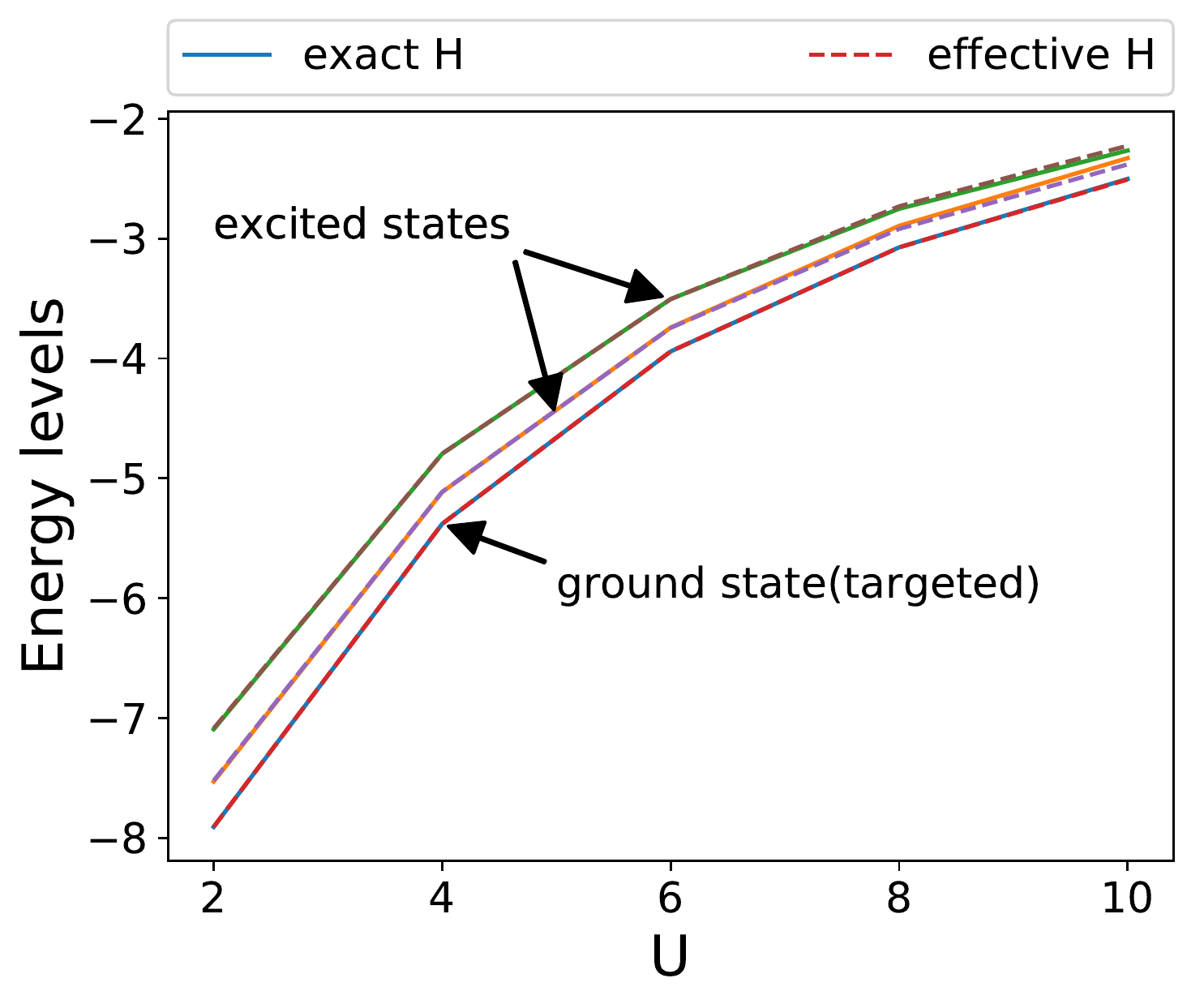} % Example image
\end{center}
\caption{The comparison between low energy levels of Effective Model yielded by the unitary and the original Hubbard model versus different U.} 
\end{figure}

Figure 6 shows the center Bond dimension of the MPO for effective Hamiltonian versus U and cutoffs for the t-J doped case with System size(N) = 6, Number of particles(P) = 4. Here we can see that doping case requires a very high bond dimension. The $10^{-2}$ cutoff is not enough to get enough accuracy, but we already need pretty big bond dimension. For the doping case, since from Hubbard to t-J model. The coefficients for high order terms are still pretty big compared to low order terms. So it is hard to yield an effective low-energy Hamiltonian with only a small number of low order terms.

\section{Conclusions}
This mapping method starting from optimizing the norm between ground states and projected ground state provides another approach to analyze effective low-energy models of strongly correlated electron systems. It works well on the half-filling case. The numerical result is consistent with the theoretical equation.  For the doping case, we can also successfully construct an effective model using ground state transform. While the MPO for effective low-energy Hamiltonian requires larger bond dimension to achieve the same accuracy compared with the half-filling case.

\begin{figure}
\centering
\begin{center} 
\includegraphics[width=0.5\columnwidth]{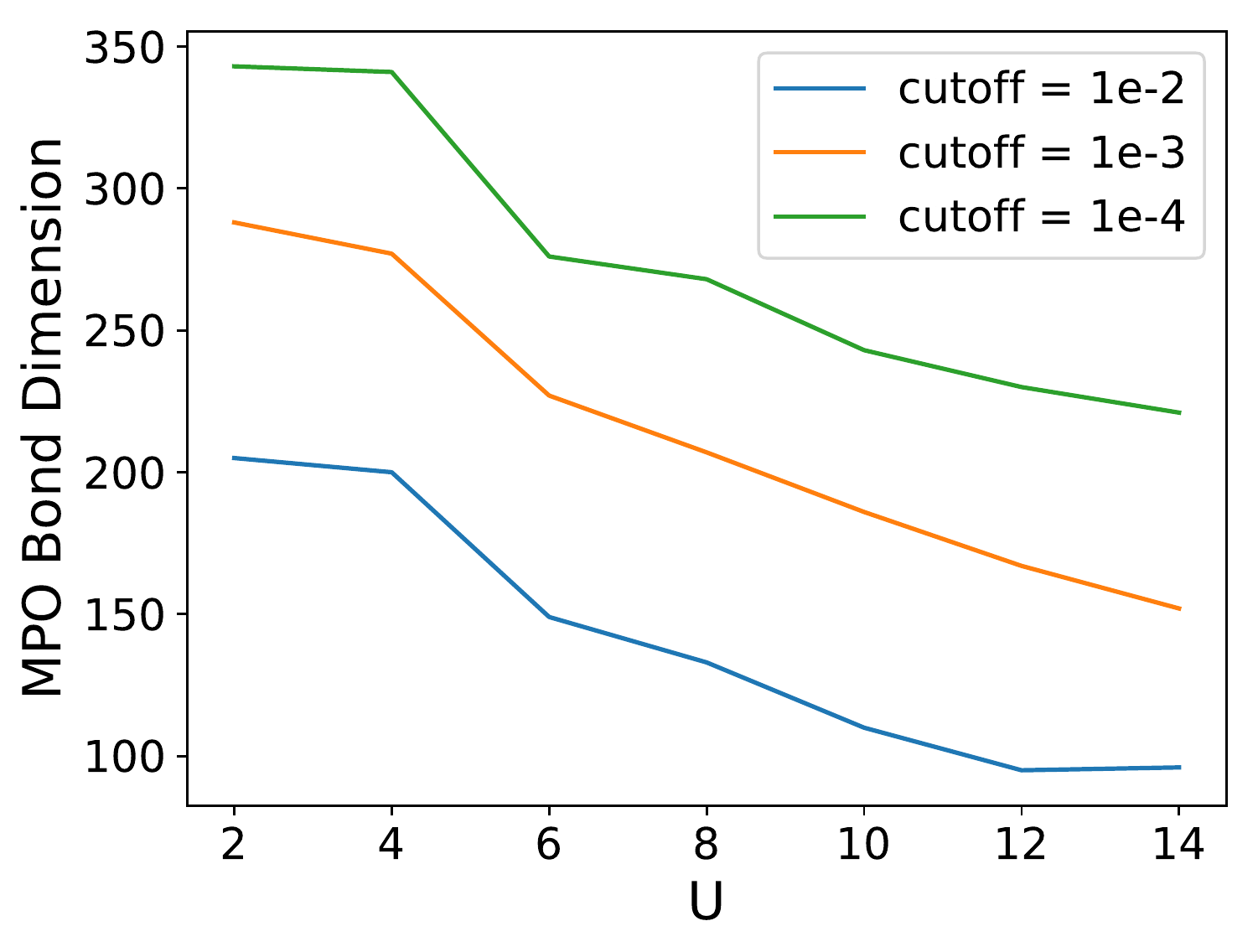} % Example image
\end{center}
\caption{The center Bond dimension of the MPO for effective Hamiltonian versus U and cutoffs for the t-J doped case with System size(N) = 6, Number of particles(P) = 4.} 
\end{figure}

\bibliographystyle{unsrt}  


\end{document}